\documentclass[preprint,aps,prb,showpacs,amsmath]{revtex4}
\usepackage{latexsym}
\usepackage{amssymb}
\usepackage{graphicx}
\usepackage{bm}
\usepackage[english]{babel}
\usepackage[latin1]{inputenc}
\usepackage{amsmath}
\usepackage{amsfonts}

\newcommand{\pdag}{{\phantom{\dagger}}}
\newcommand{\bq}{\begin{equation}}
\newcommand{\eq}{\end{equation}}
\newcommand{\bn}{\begin{eqnarray}}
\newcommand{\en}{\end{eqnarray}}

\begin{document}

\title{Elimination of negative differential conductance in an asymmetric molecular transistor 
by an ac-voltage}

\author{Bing Dong}
\affiliation{Department of Physics, Shanghai Jiaotong University,
1954 Huashan Road, Shanghai 200030, China}

\author{X.L. Lei}
\affiliation{Department of Physics, Shanghai Jiaotong University,
1954 Huashan Road, Shanghai 200030, China}

\author{N. J. M. Horing}
\affiliation{Department of Physics and Engineering Physics, Stevens
Institute of Technology, Hoboken, New Jersey 07030, USA}

\begin{abstract}

We analyze resonant tunneling subject to a non-adiabatic time-dependent bias-voltage through 
an asymmetric single molecular quantum dot with coupling between the electronic and 
vibrational degrees of freedom using a {\em Tien-Gordon-type} rate equation. 
Our results clearly exhibit the appearance of photon-assisted satellites in the 
current-voltage characteristics and the elimination of hot-phonon-induced negative 
differential conductance with increasing ac driving amplitude for an asymmetric system. This 
can be ascribed to an {\em ac-induced suppression} of unequilibrated (hot) phonons in an 
asymmetric system.
 
\end{abstract}

\date{\today}


\maketitle

Recently, phonon-mediated dc tunneling through a quantum dot (QD) with coupling to an 
internal vibrational (phonon) mode (IVM) has attracted much interest, both 
experimental\cite{hPark,LeRoy,Sapmaz} and 
theoretical.\cite{Bose,Alexandrov,Mitra,Koch,Nowack,Koch2,Zazunov,Shen} 
Electronic transport measurements in suspended Carbon nanotubes (CNT) indicate that the 
current-voltage characteristics display perfect signatures of phonon-mediated tunneling, e.g. 
stepwise structures having equal widths in voltage and gradual height reduction by the 
Franck-Condon (FC) factor.\cite{LeRoy,Sapmaz} Furthermore, striking negative differential 
conductance (NDC) and current peaks have been observed at the onsets of phonon 
steps,\cite{Sapmaz} which are ascribed to the combination of highly asymmetric 
tunnel-coupling rates and the voltage-triggered unequilibrated (hot) phonon 
effect.\cite{Koch2,Zazunov,Shen} On the other hand, ac-transport through a single molecule 
has also attracted considerable interest, but without consideration of electron-phonon 
coupling (EPC).\cite{actransport}    
In this letter, we will examine ac-transport in the presence of EPC when a time-dependent 
bias-voltage is applied between the two electrodes in the non-adiabatic regime.\cite{Dong} We 
find that application of a strong ac-amplitude causes suppression of hot phonons and thus 
elimination of NDC for a large asymmetric system.   

We consider a quantum dot (QD) with one spinless level coupled to two electrodes, and also 
linearly coupled to a molecular IVM. The model Hamiltonian is
\begin{subequations}
\bq
H = H_{leads}+H_{mol}+H_B+H_{T}, \label{hamiltonian}
\eq
with
\bn
H_{leads} &=& \sum_{\eta, {\bf k}} \varepsilon_{\eta {\bf k}} c_{\eta {\bf k}}^\dagger 
c_{\eta {\bf k}}^\pdag, \\
H_{mol} &=& \varepsilon_d c_d^{\dagger} c_d^\pdag + \omega_0 a^\dagger a + \lambda 
c_d^\dagger c_d^\pdag (a^\dagger + a), \label{Hmol} \\
H_T &=& \sum_{\eta,{\bf k}} (V_{\eta} c_{\eta {\bf k}}^\dagger c_d + {\rm H.c.}),
\en
\end{subequations}
where $c_{\eta{\bf k}}^\dagger$ ($c_{\eta{\bf k}}$) is the creation (annihilation) operator 
of an electron with momentum ${\bf k}$ in lead $\eta$ ($\eta=L,R$). 
The energies $\varepsilon _{\eta \mathbf{k}}(t)= \varepsilon _{\eta 
\mathbf{k}}^0+eU_{\eta}(t)$ include rigid shifts of the Fermi energies of the electrons in 
the leads due to the applied time-dependent bias-voltage, $U_{\eta}(t)=U_{\eta}^0 + 
u_{\eta}\cos(\Omega t)$, with $U_{\eta}^0$ ($u_{\eta}$) being the dc(ac) part of bias-voltage 
and $\Omega$ is the driving frequency. $c_{d}^\dagger$($c_d$) is the creation (annihilation ) 
operator for a spinless electron in the QD. $a^\dagger$($a$) is the phonon creation 
(annihilation) operator for the IVM with energy $\omega_0$. $\lambda$ represents the EPC 
constant; $V_{\eta}$ describes the tunnel-coupling between the QD and lead $\eta$.
Here, we assume that the Fermi energies of two leads are zero at equilibrium and 
$U_{L}^0=-U_{R}^0=eV/2$, $u_L=-u_R=eV_{ac}$ ($V$ and $V_{ac}$ are the dc bias-voltage and 
ac-bias amplitudes, respectively).   
In the following, we will use units where $\hbar=k_{B}=e=1$.

It is well-known that the electron-phonon interaction term in Eq.~(\ref{Hmol}) can be 
eliminated by a canonical transformation,\cite{Mahan} leading to a renormalization of the 
parameters, $\widetilde{\varepsilon}_d=\varepsilon_d-g\lambda$ ($g=\lambda/\omega_0$), and of 
the tunnel-coupling, $V_\eta \exp [g(a^\dagger +a)]$. In the weak tunneling regime and high 
temperature approximation, $\Gamma_\eta \ll T, \omega_0$ ($\Gamma_\eta$ is the tunneling rate 
of lead $\eta$ and $T$ is the temperature), rate equations for the electron-phonon joint 
probabilities, $\rho_{00}^n$ and $\rho_{11}^n$ for zero- and one-electron together with $n$ 
excited phonons on the molecule (incorporated with the FC-modified tunneling rates) are 
physically appropriate for the description of resonant tunneling through a single molecule 
involving EPC.\cite{Mitra,Koch,Nowack,Koch2,Zazunov,Shen} On the other hand, 
in the limit of high driving frequency, $\Omega\gg \Gamma, T$, of interest in this letter, 
the ac-bias oscillates so fast that an electron experiences many cycles of the ac-bias during 
its presence inside the dot, and thus can not sense the details of the dynamics within a 
single period. Correspondingly, the rate equations can be established by directly applying a 
Tien-Gordon-type tunneling rate in the presence of such an ac-bias in the non-adiabatic 
limit,\cite{Tien,Wiel}
\bn
\dot \rho_{00}^n &=& \sum_{m} ( \Gamma_{nm}^- \rho_{11}^m - \Gamma_{nm}^+ \rho_{00}^n ), 
\label{r00} \\
\dot \rho_{11}^n &=& \sum_{m} (\Gamma_{mn}^+ \rho_{00}^m - \Gamma_{mn}^- \rho_{11}^n ), 
\label{r11} 
\en
with the normalization relation $\sum_{n} (\rho_{00}^n +\rho_{11}^n) =1$. The electronic 
tunneling rates are defined as
\bn
\Gamma_{nm}^+ &=& \sum_{\eta} \Gamma_{\eta ,nm}^+ = \sum_{\eta} \Gamma_{\eta } \gamma_{nm} 
\sum_{j=-\infty}^{\infty} \left [ J_{j} \left ( \frac{u_{\eta}}{\Omega} \right ) \right ]^2 
\cr
&& \times f_{\eta}(\widetilde {\epsilon}_d + (m-n)\omega_0-j\Omega),  \label{g1} \\
\Gamma_{nm}^- &=& \sum_{\eta} \Gamma_{\eta ,nm}^- = \sum_{\eta} \Gamma_{\eta } \gamma_{nm} 
\sum_{j=-\infty}^{\infty} \left [ J_{j} \left (\frac{u_{\eta}}{\Omega}\right )\right ]^2 \cr
&& \times [1-f_{\eta}(\widetilde {\epsilon}_d + (m-n)\omega_0-j\Omega)], \label{g2}
\en
with the {\em FC factor} given by ($p={\rm min}\{m,n\}$ and $q={\rm max}\{m,n\}$, denoting 
the smaller and larger of the quantities $m$ and $n$, respectively)
\bq
\gamma_{nm} = e^{-g^2} g^{2|m-n|} \frac{p!}{q!}\bigl[ L_{p}^{|m-n|}(g^2)\bigr]^2,
\eq
where $L_n^m(x)$ is the generalized Laguerre polynomial.
$J_j(x)$ is the Bessel function of order $j$. $f_{\eta}(\epsilon)=[1+ 
e^{(\epsilon-\mu_{\eta})/T} ]^{-1}$ is the Fermi-distribution function. 
Moreover, the dc current can be calculated by solving the rate equations, Eqs.~(\ref{r00}) 
and (\ref{r11}), in the steady state condition and evaluating the net tunneling rate through 
one of electrodes (for example, the left lead):
\bq
I = \sum_{nm} \bigl( \Gamma_{L,nm}^+ \rho_{00}^n - \Gamma_{L,mn}^- \rho_{11}^n \bigr).  
\label{currentL}
\eq                     
Note that if the driving amplitude vanishes, $u_\eta=0$, the above equations 
(\ref{r00})-(\ref{currentL}) reduce exactly to the rate equations formulated without an 
ac-bias.\cite{Mitra,Koch,Nowack,Koch2,Zazunov,Shen}

In our numerical calculation, we set the parameters as: $\omega_0=1$ is the energy unit, 
$\Gamma_{R}/\Gamma_{L}=10^3$, $g=1$, $\widetilde{\varepsilon}_d=0$, $T=0.02\omega_0$, and the 
driving frequency is $\Omega=0.5\omega_0$. Note that we choose a large asymmetry in the 
tunneling rates, $\Gamma_{R}/\Gamma_{L}=10^3$, and an intermediate EPC strength, $g$. The 
transport properties of this specific system exhibit weak NDC at the onsets of phonon steps 
under hot phonon condition.\cite{Zazunov,Shen} In particular, numerical fits of the 
experimentally measured data for the $I$-$V$ curves show $g\simeq 1$ for long 
CNTs.\cite{Sapmaz}

Figure 1 shows the dc current-bias-voltage characteristics as a function of various driving 
amplitudes. For comparison, we also plot the results without an ac-bias, $V_{ac}=0$, as the 
solid line, showing (1) step behavior with width equal to $2\omega_0$, (2) the FC 
factor-suppressed height, and (3) NDC at the onsets of phonon steps starting from the second 
step. When an ac-bias is applied, it is obvious to see from Fig.~1 that a {\em 
photon-assisted} tunneling feature---constituted of new steps at $V=2n\Omega$ ($n$ is 
integer)---is superimposed on the phonon-mediated $I$-$V$ characteristics. For stronger 
driving amplitudes, the suppression of current is more pronounced in the weak dc-bias region, 
$V<2\Omega$. However, in the strong dc-bias region, $V>4\omega_0$, the application of the 
ac-bias always enhances the current. Such nonuniform behavior is reflected in the structure 
of the Bessel function in the tunneling rates, Eqs.~(\ref{g1}) and (\ref{g2}).
More interestingly, we observe that for a weak ac-amplitude, $V_{ac}=0.2\omega_0$, the 
position of NDC moves to the ends of the phonon steps, while in the cases of strong 
ac-amplitudes, $V_{ac}=0.5\omega_0$ and $1.0\omega_0$, the NDC even {\em disappears}. It 
should be noted that strong vibrational relaxation can result in the elimination of NDC even 
in the absence of an ac-bias.\cite{Zazunov,Shen} In that case, photon-assisted steps still 
remain in the current-voltage characteristics when an ac-bias is applied (not shown 
here).\cite{Dong} 
 

To gain insight into this peculiar behavior, we adopt the method of Ref.~\onlinecite{Zazunov} 
by examining the phonon occupation numbers (PONs), $\rho_{00}^n$. For the positive dc 
bias-voltage range of interest in this letter, $V>0$, the QD is always in an empty occupation 
state owing to $\Gamma_{R}\gg \Gamma_L$ (so that the electron has a much stronger 
tunneling-out rate to the right lead than the tunneling-in rate from the left lead). 
Therefore, a nonzero value of $\rho_{00}^n$ indicates an open channel contributing to the 
current. In this case, we can rewrite the current formula, Eq.~(\ref{currentL}), with good 
accuracy, as
\bq
I = \sum_{n}^N {\cal I}_n \rho_{00}^n ,  \label{current}
\eq
with ($N$ denotes the maximum number of PONs of the open channels)
\bn                     
{\cal I}_n &=& \sum_m \Gamma_{L,nm}^+ = \Gamma_L \sum_{m} \gamma_{nm} 
\sum_{j=-\infty}^{\infty} \left [ J_{j} \left ( \frac{u_{\eta}}{\Omega} \right ) \right ]^2 
\cr
&& \times \Theta(-\widetilde {\epsilon}_d - (m-n)\omega_0+j\Omega + V/2) \label{In} 
\en
[$\Theta(x)$ is the Heaviside step function]. Starting from these two equations, we interpret 
the NDC and positive differential conductance (PDC) behaviors of the $I$-$V$ characteristics 
in the presence of the ac-bias in Fig.~1.
 

Figure 2 displays the $\rho_{00}^n$ under several dc-bias voltages and ac-amplitudes as well.
It is evident that for small dc-bias, $V_1=2.0\omega_0$ (the panels of the first row in 
Fig.~2), only the channel with the PON $n=0$ is active, $\rho_{00}^0\approx 1$, and the 
application of an ac-bias has very little influence on the PON. In these situations, we have 
$N=0$ and thus
\bq
I_1\simeq \Gamma_L \sum_{m} \gamma_{0m} {\cal J}_m, 
\eq
with ${\cal J}_m=\sum_{j=2(m-1)}^\infty [ J_{j} ( \frac{V_{ac}}{\Omega} ) ]^2$.
If the ac-bias vanishes, the current becomes (we use $\bar I$ here to denote the current in 
the absence of an ac-bias) 
\bq
\bar I_1\simeq \Gamma_L (\gamma_{00}+ \gamma_{01}).
\eq
When the dc-bias exceeds the transition point $V_1=2.0\omega_0$, for example 
$V_2=2.3\omega_0$ (the panels in the second row of Fig.~2), two channels, $n=0$ and $1$, are 
open in the absence of an ac-bias, which indicates an {\em unequilibrated (hot) phonon} 
distribution. Correspondingly, the current is
\bn
\bar I_2 &\simeq& \Gamma_L [ \rho_{00}^0 (\gamma_{00} + \gamma_{01}) + \rho_{10}^1 
(\gamma_{10} + \gamma_{11} + \gamma_{12})] \cr
&\simeq& \Gamma_L \gamma_{00} + \Gamma_L \rho_{00}^1 (\gamma_{11} + \gamma_{12}- 
\gamma_{00}),
\en
because $\rho_{00}^0 + \rho_{00}^1\simeq 1$. As a result, $\bar I_2- \bar I_1\simeq \Gamma_L 
\rho_{00}^1 e^{-g^2} \rho_{00}^1 g^4 (g^2/2-1)<0$ due to $g=1$ in this letter, leading to the 
appearance of NDC.\cite{Zazunov}
When an ac-bias is applied, we find---strikingly---from Fig.~2 that the hot phonon is 
suppressed and the $n=1$ channel becomes closed. This suppression of $\rho_{00}^1$ stems from 
a weak increase of $\rho_{11}^n$ due to photon-assisted processes. As a result, we have 
$I_2\approx I_1$ and the NDC reverts to a pure PDC at the onsets of the phonon steps. 
Furthermore, for $V_{3}=3.0\omega_0$ and $V_4=3.3\omega_0$, two channels, $n=0$ and $1$, are 
activated by the dc-bias and we readily obtain $\bar I_3\simeq \bar I_4$. The ac-bias-induced 
suppression of hot phonons changes the current of the case of $V_3$ as $I_3\simeq I_1$. 
However, the application of a weak ac-amplitude, $V_{ac}=0.2\omega_0$, can not completely 
suppress the $n=1$ hot phonon at the moderately stronger dc-bias $V_4$, and thus
\bn
I_4 &\simeq& \Gamma_L \rho_{00}^0 \sum_{m} \gamma_{0m} {\cal J}_m + \Gamma_L \rho_{00}^1 
\sum_{m} \gamma_{1m} {\cal J}_m' \cr
&=& I_3 + \Gamma_L \rho_{00}^1 \sum_{m} (\gamma_{1m} {\cal J}_m' - \gamma_{0m} {\cal J}_m), 
\label{i4} 
\en 
with ${\cal J}_m'=\sum_{j=2m-5}^\infty [ J_{j} ( \frac{V_{ac}}{\Omega} ) ]^2$. The second 
term of the last line of Eq.~(\ref{i4}) may be positive or negative depending on the EPC 
constant $g$ and the ac-amplitude $V_{ac}$. For the weak ac-amplitude $V_{ac}=0.2\omega_0$, 
it is negative, leading to NDC; while for stronger amplitudes, $V_{ac}=0.5\omega_0$ and 
$1.0\omega_0$, it is positive and PDC is observed. 

In summary, we have employed the Tien-Gordon-type rate equations incorporated with the FC 
model to analyze photon-assisted resonant tunneling through an asymmetric single molecular QD 
with EPC in the non-adiabatic and high frequency regime. Our results predict that the 
application of an ac-bias causes {\em suppression} of the dc-bias-triggered hot phonon, 
leading to the {\em elimination} of NDC in the $I$-$V$ characteristics.

This work was supported by Projects of the National Science Foundation of China, the Shanghai 
Municipal Commission of Science and Technology, the Shanghai Pujiang Program, and Program for 
New Century Excellent Talents in University (NCET).

\newpage

\centerline{\Large Figure Caption}

\vspace{2cm}

\noindent FIG.1: $I$-$V$ curves as functions of ac bias-voltage amplitudes, 
$V_{ac}/\omega_0=0$, $0.2$, $0.5$, and $1.0$, for a fixed driving frequency, 
$\Omega/\omega_0=0.5$. The parameters we use in the calculation are: $g=1$, 
$\Gamma_L/\Gamma_R=10^{-3}$, and $T/\omega_0=0.02$.

\vspace{1cm} \noindent FIG.2: Phonon probability distributions for increasing ac-amplitude, 
$V_{ac}$, under different dc bias-voltages, $V/\omega_0=2.0$, $2.3$, $3.0$, and $3.3$. Other 
parameters are the same as in Fig.~1.

\newpage

\begin{figure}[htb]
\includegraphics [width=8.5cm,height=6cm,angle=0,clip=on]{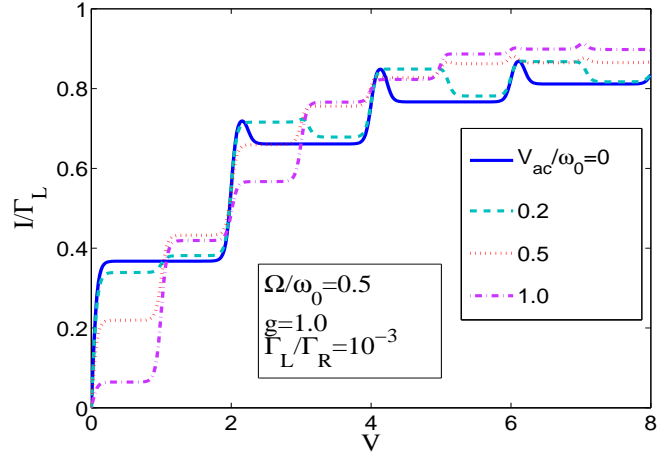}
\caption{$I$-$V$ curves as functions of ac bias-voltage amplitudes, $V_{ac}/\omega_0=0$, 
$0.2$, $0.5$, and $1.0$, for a fixed driving frequency, $\Omega/\omega_0=0.5$. The parameters 
we use in the calculation are: $g=1$, $\Gamma_L/\Gamma_R=10^{-3}$, and $T/\omega_0=0.02$.} 
\label{fig1}
\end{figure}

\newpage

\begin{figure}[htb]
\includegraphics [width=8.5cm,height=9.5cm,angle=0,clip=on]{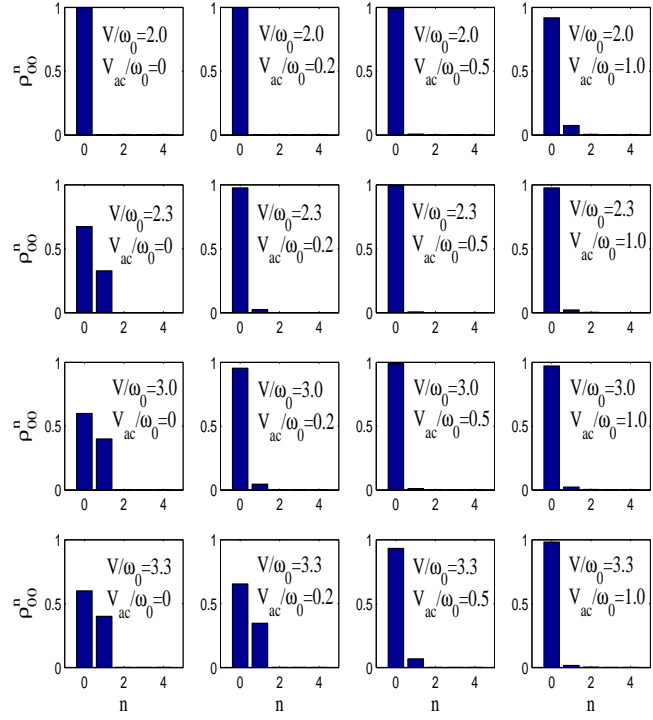}
\caption{Phonon probability distributions for increasing ac-amplitude, $V_{ac}$, under 
different dc bias-voltages, $V/\omega_0=2.0$, $2.3$, $3.0$, and $3.3$. Other parameters are 
the same as in Fig.~1.} \label{fig2}
\end{figure}

\end{document}